\documentclass[12pt]{article}
\usepackage{sao1}
\usepackage{psfig}

\begin{document}

\title{
The ZOO of $uvby$~and $H_{\rm p}$~light curves of magnetic
chemically peculiar stars }

\author{
Mikul\'{a}\v{s}ek Z.\inst{1} \and Zverko J.\inst{2} \and Krti\v{c}ka
J.\inst{1} \and Jan\'\i k J.\inst{2} \and \v{Z}i\v{z}\v{n}ovsk\'{y}
J.\inst{2} \and Zejda M.\inst{1} }

\institute{ Institute of Theoretical Physics and Astrophysics,
    Kotl\'{a}\v{r}sk\'{a} 2, CZ-637 00 Brno, Czech Republic \and
    Astronomical Institute, Slovak Academy of Sciences,
    SK-059 60 Tatransk\'{a} Lomnica, Slovak Republic}

\maketitle

\begin{abstract}
We present preliminary results of the generalized Principal
Component Analysis (PCA) of light curves of 82 magnetic chemically
peculiar (further mCP) stars applied to 54 thousand individual
photometric observations in the \textit{uvby} and $H_{\rm
p}$~colours taken from the ``On-line database of photometric
observations of mCP stars".

We show that every of the observed light curves (LCs) can be, with a
sufficient accuracy, represented using five parameters of a harmonic
polynomial of the second order, and that the third order and higher
harmonics reflect only a noise. We found that a prevailing majority
of the $uvbyH_{\rm p}$ LCs can be satisfactorily well represented by
a linear combination of a constant term and one or two
(extraordinarily three) basic, mutually orthonormal functions, which
reduces the number of parameters necessarily needed to describe the
set of the LCs of a particular star.

While the shape of an individual LC depends on the individual
distribution of photometric spots on the surface of a rotating star,
its amplitude depends on the mechanism of variability. To describe
the amplitude of LCs we introduce a robust quantity {\it effective
amplitude}. Applying the PCA to the set of the effective amplitudes
of the LCs in all five colours we revealed at least three different
sources of the light variability showing different wavelength
dependence. The amplitudes of all the LCs harbor a component
monotonously decreasing with increasing wavelengths. The second
component reaches a remarkable extremum in the $v$ colour, and is in
antiphase relative to the remaining four colours. It occurs
particularly at cooler mCP-s having a high degree of the chemical
peculiarity. The third component reflects a remarkable diversity of
the light curves in the $u$ colour found in a few mCP stars.

\end{abstract}

\section{Introduction}
The light variability of the magnetic chemically peculiar (mCP)
stars, ranging a few hundredths, seldom a few tenths of magnitude in
amplitude, has been known and investigated for decades. These
strictly periodic light variations, frequently accompanied by
synchronous variations of spectral lines and magnetic field, are
commonly explained due to existence of hypothetical, vast and
persistent \textit{photometric spots} on the surface of rigidly
rotating stars. The properties of these photometric spots and their
relation to the known spectroscopic spots and the geometry of the
magnetic field have not been explored satisfactorily up to now.

We aim at a deep analysis of the rotationally
modulated photometric variations of mCP stars, their astrophysical
explanation based both on the complex study of selected well
observed mCP stars and the study of the photometric behaviour of as
large as possible sample of the stars.

\section{The data}
Prior to start the investigation we established a publicly
accessible ``On-line database of photometric observations of mCP
stars" (\emph{http://dumbell.physics.muni.cz/pilot}, hereafter the
Database) as today containing more than 120 thousand observations of
more then 120 mCP stars made in 40 photometric colours. One can find
a more detailed information on this public database in the paper
\cite{mikdat}.

We selected 82 stars having observations in $u, v, b, y$ and
Hipparcos $H_{\rm p}$ colours. The sample contains one He-strong
(HD\,37776), 10 He-weak, 37 Si and 34 ``cool'' SrCrRE stars. Here we
study altogether 408 LCs in five photometric colours consisting of
54\,131 individual observations with errors of $\approx 0.003$\,mag.

The linear light elements of all the 82 stars were determined by
means of our own method based on the generalized PCA and a robust
regression, based on the complete photometric material contained in
the Database. The fastest rotator in our sample is HD\,164429 with
the rotational period $P=0.519$\,d, the slowest one is HD\,188041
with $P=224$\,d. The median is $P=3.8$\,d.

\begin{figure}
\centerline{\psfig{figure=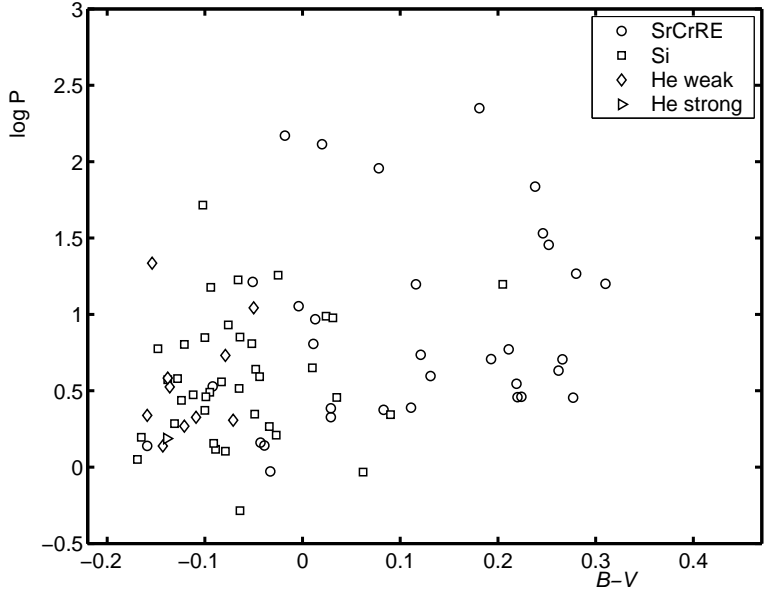,height=8.cm}} \caption{Dependence
of the rotational period on the mean Hipparcos \emph{B-V} index.}
\label{bvp}
\end{figure}

An evident trend that the cooler mCP stars rotate slower on average,
can be seen on Fig. \ref{bvp}. The undereddened \emph{B-V} indices
were taken from the Hipparcos photometry.

\section{Description of mCP light curves}

The rotational LCs of magnetic CP stars are strictly periodic what
means they can be described to advantage by harmonic polynomials
depending on the rotational phase $\varphi$. Further, it follows
from the experience (\cite{north}) that the rotational LCs are
smooth to such extent that the polynomial of the 2nd degree is quite
sufficient. The terms of the higher degrees mostly reflect noise so
they can be neglected. Thus
\begin{equation}
m(c,\varphi)=\overline{m_c}+a_{c1}\cos(2\pi\varphi)+a_{c2}\sin(2\pi
\varphi)+a_{c3}\cos(4\pi\varphi)+a_{c4}\sin(4\pi\varphi)=\overline{m_c}
+\mathbf{a_\mathit{c}}\cdot\,\mathbf{CS}(\varphi), \label{Eq1}
\end{equation}
where the dot ``$\cdot$'' denotes a scalar product. The fourvector
$\mathbf{CS}(\varphi)$ is given by the relation:
\begin{equation}
\mathbf{CS}(\varphi)=[\cos(2\pi\varphi),\ \sin(2\pi \varphi),\
\cos(4\pi\varphi),\ \sin(4\pi\varphi)]. \label{Eq2}
\end{equation}
The shape of the light curve in the the colour $c$ is then described
simply by a fourvector $\mathbf{a_\mathit{c}}$,
$\mathbf{a_\mathit{c}}=[a_{c1},a_{c2},a_{c3},a_{c4}]$, consequently
by a point in the 4D-space of parameters. The shapes of the set of
$uvbyH_{\rm p}$ light curves of each of the 82 mCP stars are fully
described by the set of 5 fourvectors which can be considered as a
$5 \times 4$ matrix \textbf{Y} with 20 elements known with different
degree of uncertainty. It is rather surprising that there are no two
stars with same or even resembling set of these vectors. Thus the
set of parameters of a given star seem to be unique and
characteristic.

\subsection{PCA decomposition of light curves of individual stars}

The thorough inspection of $uvbyH_{\rm p}$ light curves of
individual stars suggests that these LCs are in some extent similar,
what entitles us to apply the PCA to them with the aim of reducing
the number of parameters needed to describe the LCs of a particular
star.

The technique of weighted PCA enables one
to search for mutual relations of five fourvectors of light
variations $\textbf{a}_c$ in the 4D-space for individual stars. We
are looking for the basic direction towards which the sum of squared
projections is maximum.

The found orthonormal base vectors
$\textbf{d}_1,\,\textbf{d}_2,\,\textbf{d}_3,\,\textbf{d}_4$
determine the course of normalized basic LCs $f_{1}(\varphi)$,
$f_{2}(\varphi)$, $f_{3}(\varphi)$, $f_{4}(\varphi)$ as follows:
\begin{equation}
f_j(\varphi)=\textbf{d}_j\cdot\mathbf{CS}(\varphi). \label{Eq3}
\end{equation}

\begin{figure}
\centerline{\psfig{figure=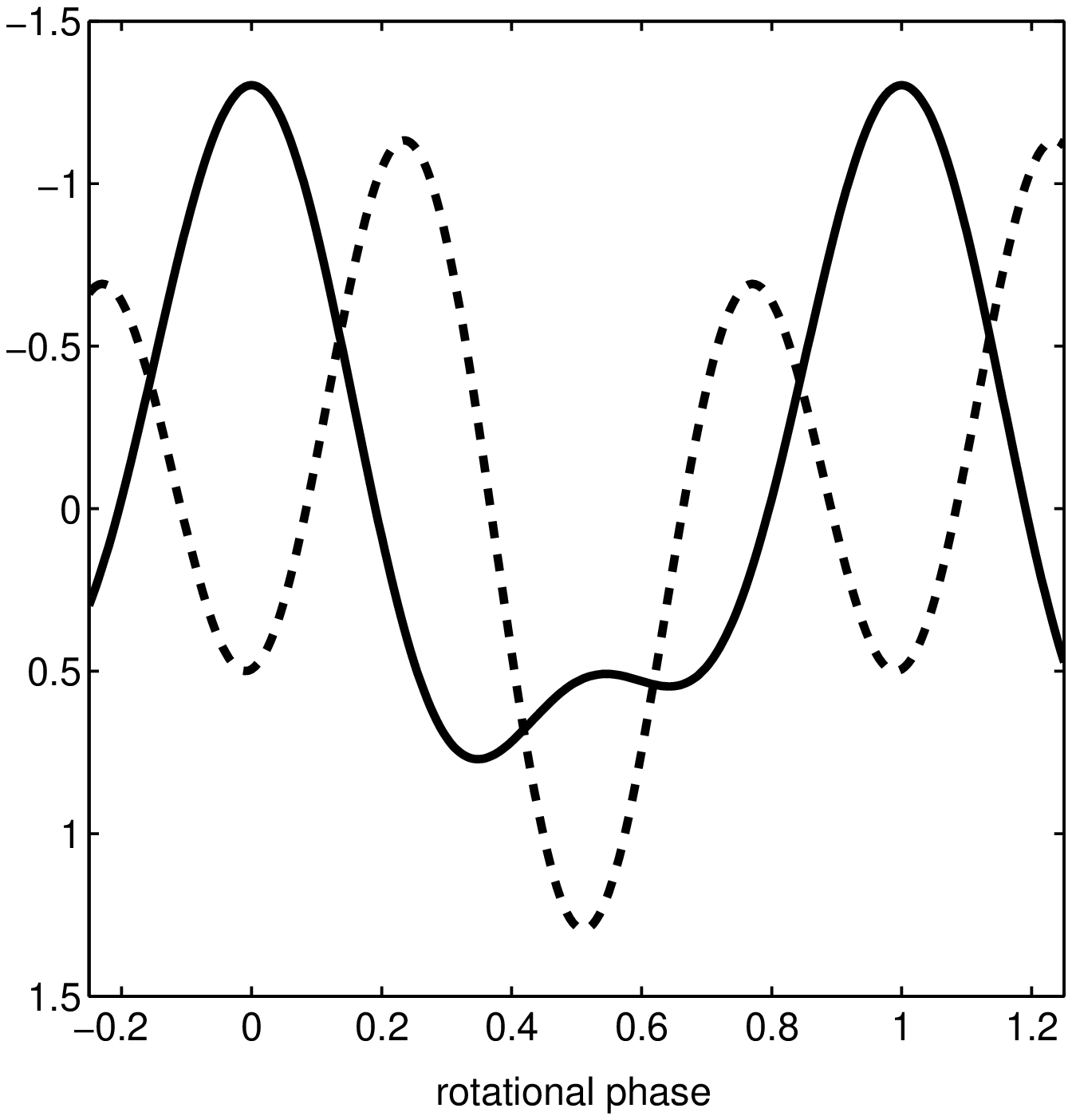,height=7.5cm}
        \psfig{figure=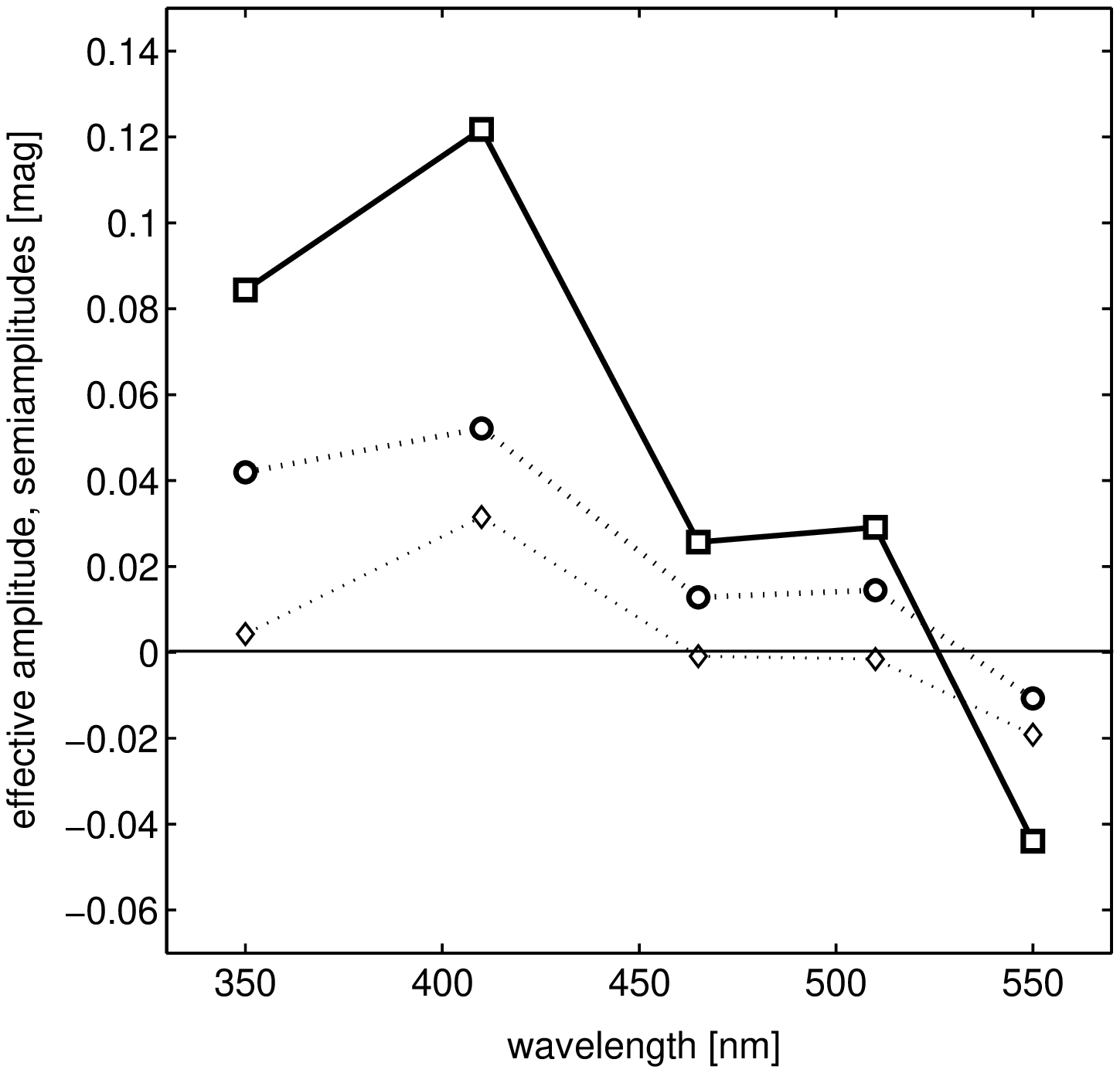,height=7.3cm}}
\caption{(a) Two basic phase functions of HD\,125248, the first one
- $f_1(\varphi)$ (full line) and the secondary one - $f_2(\varphi)$
(dotted line). (b) The dependence of semiamplitudes $A_{1c}$,
$A_{2c}$ and the effective amplitude of main and secondary
components on wavelength in nm } \label{TwoPCA}
\end{figure}

\begin{figure}
\centerline{\psfig{figure=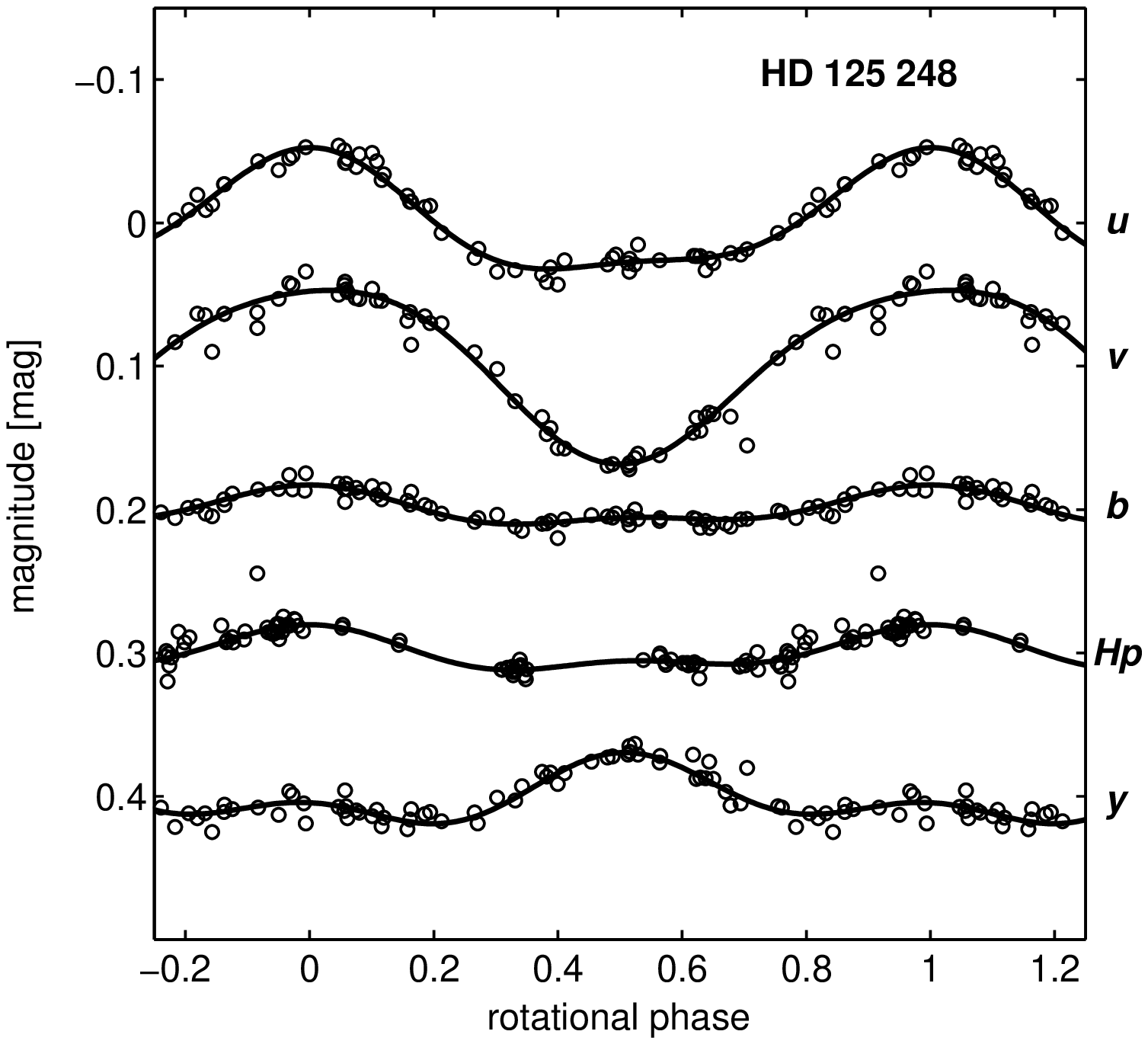,height=7.5cm}
        \psfig{figure=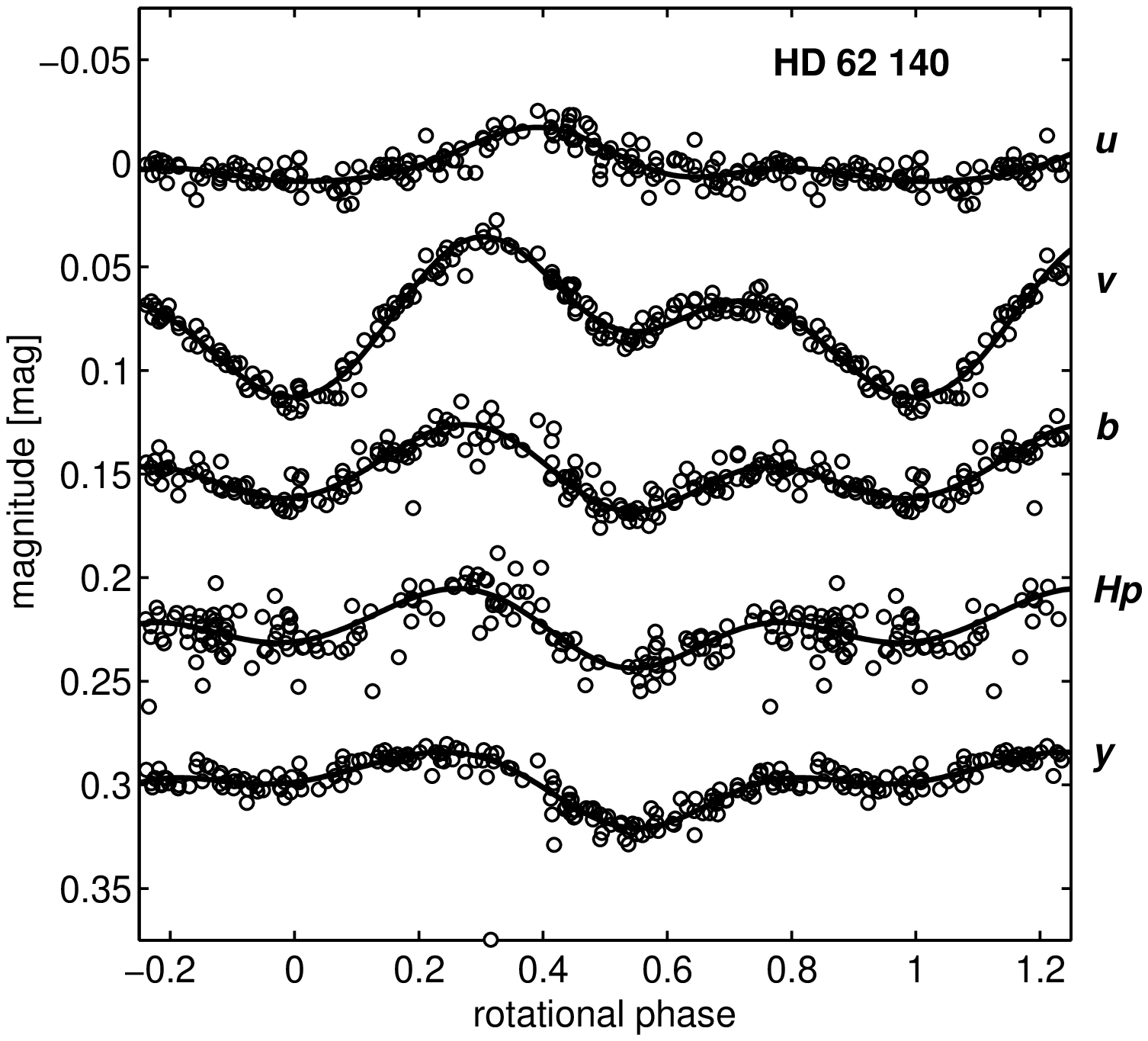,height=7.5cm}}
\caption{All the LCs of HD\,126515 can be satisfactorily expressed
by the linear combination of two basic light curves. HD\,62140 is
the only star where we need three basic light curves.}
\label{TwoLC3}
\end{figure}

We found that a colour light curve $m_c(\varphi)$ can be expressed
as a linear combination of at most the first three basic functions
$f_j(\varphi)$:
\begin{equation}
m_{c}(\varphi)=\overline{m_c}+\sum_{j=1}^kA_{jc}\,f_{j}(\varphi)=
\overline{m_c}+\left(\sum_{j=1}^kA_{jc}\
\textbf{d}_j\right)\!\cdot\mathbf{CS}(\varphi),\ \ \ k\leq
3.\label{Eq4}
\end{equation}
The first basic function, $f_1(\varphi)$ represents common features
of the course of all 5 light curves, while the second,
$f_2(\varphi)$, expresses differences from each other. The need of
the third function, $f_3(\varphi)$, occurs seldom. The accuracy of
0.001 mag of the fitting of the LCs could be achieved with only one
function $f_1$ in 34 cases, we had to use two functions $f_1$ and
$f_2$ (see Fig. \ref{TwoPCA}a) in 47 cases, and we needed three
functions $f_1,f_2$ and $f_3$ only in the case of HD\,62140 (see
Fig.~\ref{TwoLC3}b).

The semiamplitudes $A_{jc}$  describe LCs in particular colours
quantitatively and give an information on mechanisms of stellar
variability (see Fig. \ref{TwoPCA}b). They can be determined simply
as projections of the vector $\textbf{a}_c$  into the vector
${\textbf{d}_j}$ of the vector base by the relation:
$A_{jc}=\textbf{d}_j\cdot\textbf{a}_c$ or better by using LSM or a
robust regression. We have applied the latter as it suppresses the
influence of outliers and furnishes us also with the robust values
of $\overline{m_c}$ for particular colours and observers and by the
estimate of the uncertainties of the found parameters.

\section{Effective amplitude of a light curve}

The effective amplitude is a robust quantity characterizing the
amplitude of a light curve. \cite{mikdat} defined the effective
amplitude $A_{\rm {eff}\it{c}}$ of a LC of an mCP star in the colour
$c$ as
\begin{equation}
\label{ampdefjed}
A_{\rm {eff}\it{c}}=\sqrt{\,8
\int_0^1(m_c({\varphi})-\overline{m_c})^2\,d{\varphi}}~~\label{Eq5}.
\end{equation}
The factor 8 is selected so that the effective amplitude of the sine
light curve corresponded to the amplitude of the observed one. Using
relations (\ref{Eq2})and (\ref{Eq4}) we arrive at relations:

\begin{equation}
A_{\rm {eff}\it{c}}=\pm2\,|\mathbf{a_\emph{c}}|,\ \ \ \ A_{\rm
{eff}\it{c}}=2\,\mathrm{sign}\left({A_{1c}}/{\sum\limits_{q=1}^5A_{1q}}
\right)\sqrt{\sum_{j=1}^kA_{jc}^2}~~. \label{Eq6}
\end{equation}

Here we extended the definition Eq.~(\ref{ampdefjed}) by introducing
the negative effective amplitude for a LC going in antiphase with
the majority of the remaining LCs. The meaning of the phrase "to be
in antiphase" is specified by the last relation in the equation
(\ref{Eq6}).

\begin{figure}
\centerline{\psfig{figure=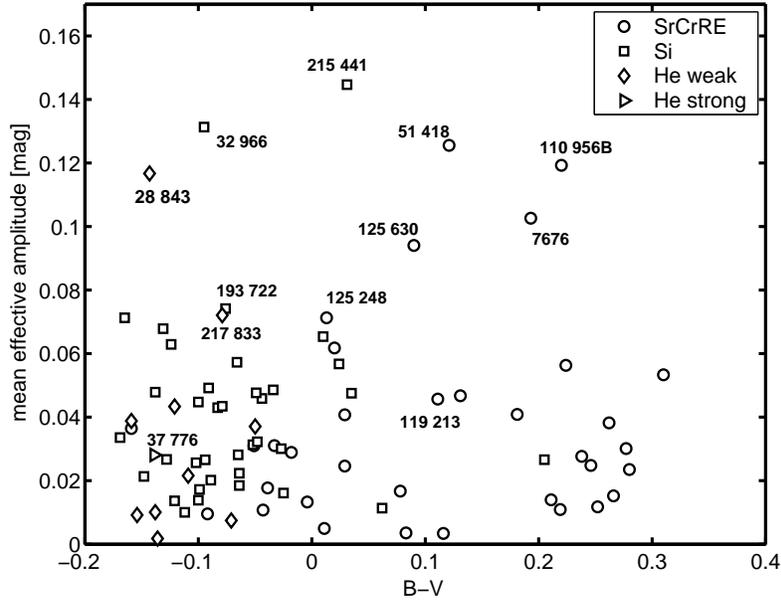,height=8.cm}} \caption{Dependence
of the mean value of the effective amplitude on the average
Hipparcos \emph{B-V}. Position of several famous mCP stars are
denoted by their HD numbers. We do not see any unambiguous relation
between the effective amplitudes and \emph{B-V} indices.}
\label{bvb}
\end{figure}

Fig. \ref{bvb} displays that there is no apparent correlation
between the mean value of effective amplitudes and \emph{B-V}
indices. We argue that mCP stars of various types form  more or less
homogeneous group from the view of their photometric variability.

\subsection{Stars with the largest effective amplitudes}

\begin{figure}
\centerline{\psfig{figure=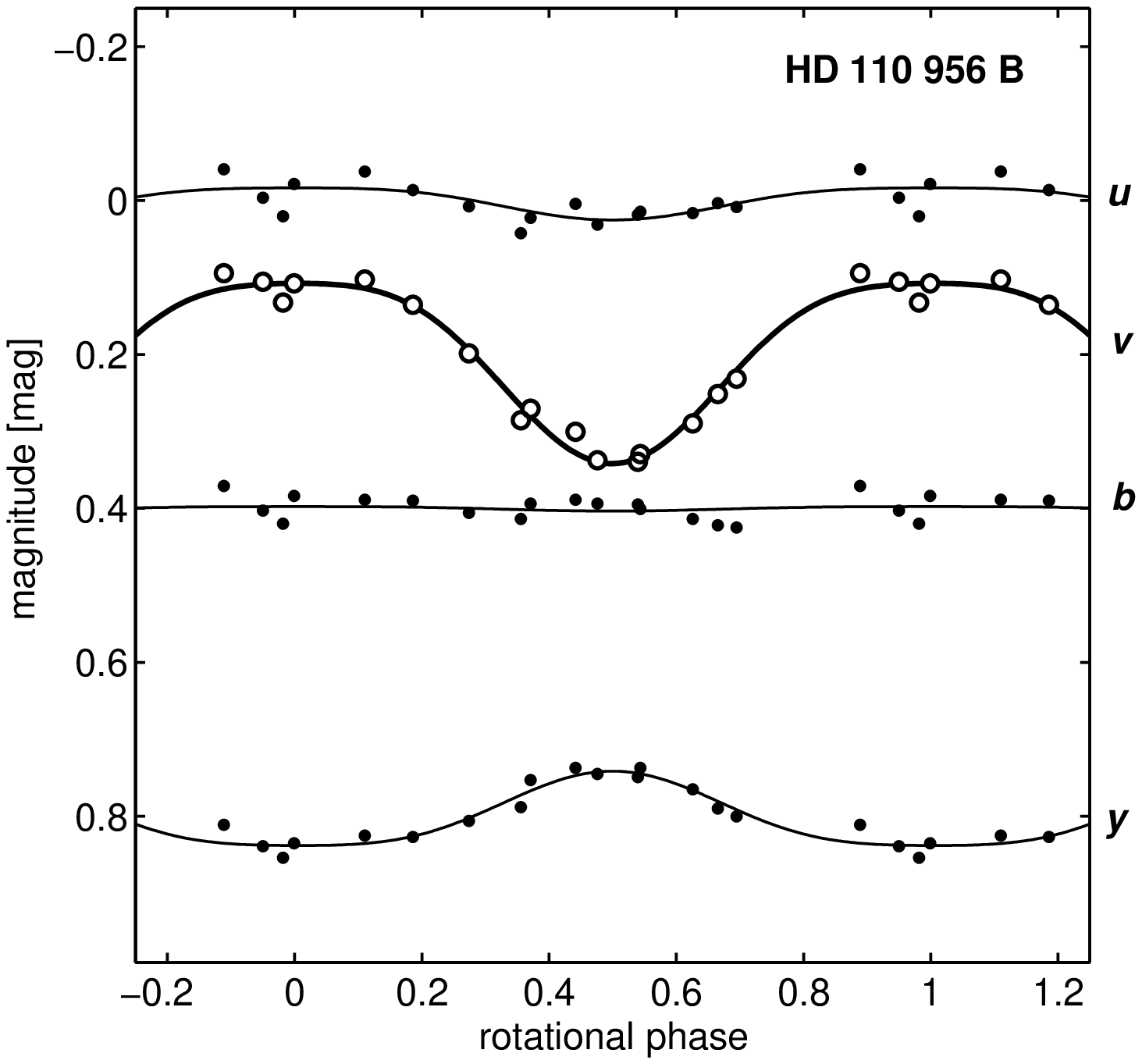,height=7.5cm}
        \psfig{figure=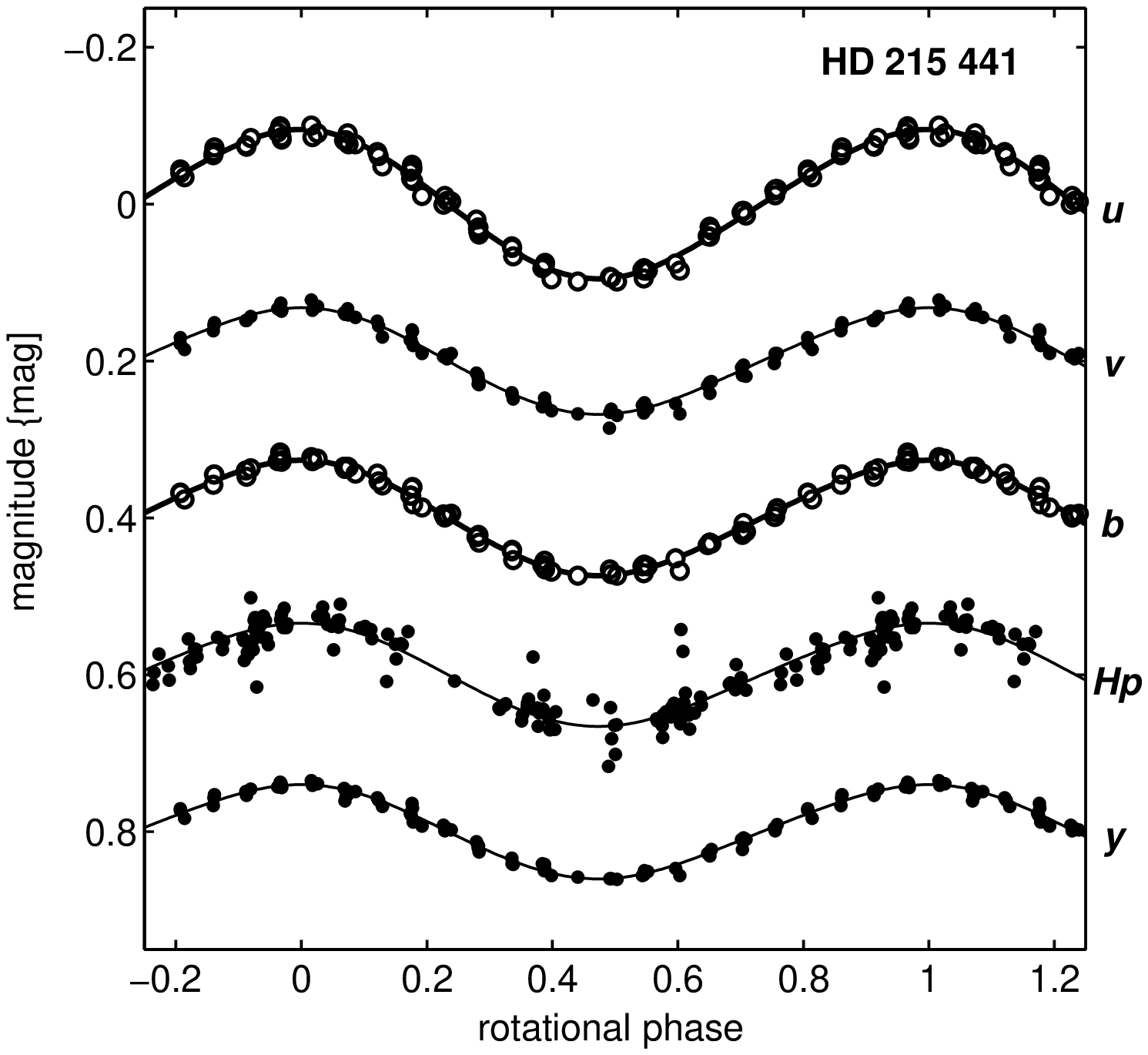,height=7.5cm}}
\caption{The comparison of LCs of two mCP stars with the largest
amplitudes. While in the case of HD\,215441 the amplitudes decrease
with the increasing wavelength, right, HD\,110956B has the largest
amplitude in $v$, and the antiphase occurs in $y$. These two stars
show two basic models of the dependence of the amplitude on the
wavelength, and represent two different mechanisms of variability of
the mCP stars.} \label{TwoLC}
\end{figure}

The median of the set of the effective amplitudes is 0.027 (!) mag.
Still some of the mCPs have much larger amplitude as documented in
the following: the Olsen's star, HD\,110956B has an effective
amplitude 0.24 mag in $v$. The Babckock's star HD\,215441 has 0.19
and 0.14 mag in $v$~and $b$~ respectively. HD\,51418 has 0.14 and
0.17 mag in $H_{\rm p}$ and $y$~respectively. HD\,28843, the most
variable He-weak star has an effective amplitude 0.15 mag in $u$.
The common characteristics of these ``record holders'' is a
single-waved LC (see Fig.\ref{TwoLC}) suggesting a simple geometry
of their photometric spots.

\subsection{PCA of the dependence of effective amplitudes on
wavelength}

The advantage of the PCA is that it offers the possibility to
investigate all the sample to be studied simultaneously. The light
variability of each star can be characterized by a fivevector of the
effective amplitudes in all five colours as $\textbf{\textit A}_{\rm
eff}=[A_{\rm {eff}\it u},A_{\rm {eff}\it v},A_{\rm {eff}\it
b},A_{\rm {eff}\it Hp},A_{\rm {eff}\it y}]$. If a LC is in
antiphase, what occurs approximately in 3.5\% of the LCs, the
corresponding effective amplitude is assigned with the negative
sign.

\begin{figure}
\centerline{\psfig{figure=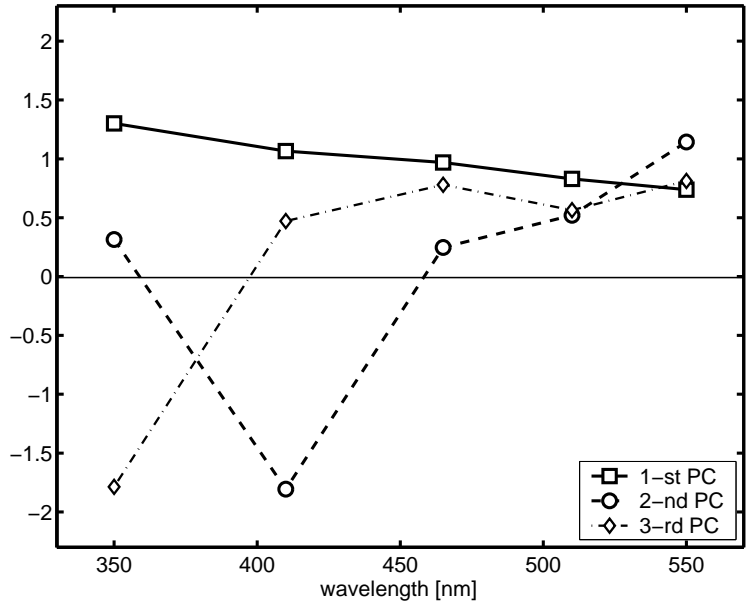,height=8.cm}} \caption{Three
most important principle components of the dependence of the
effective amplitude on the wavelength. The first component occurs in
all the sample stars and we ascribe it to a bright photometric spot
or spots. The second component occurs rather in the cooler mCP stars
and corresponds to the incidence of a spot extremely dark in the $v$
colour. We argue it is the consequence of the diminishing of the
Balmer jump in heavily contaminated atmospheres. The third component
occurs only in a few stars having anomalous appearance of the LC in
$u$.} \label{3main}
\end{figure}

Then these fivevectors $\textbf{\textit A}_{\rm eff}$ are, using the
PCA, expressed by the linear combinations of at least as possible
number of the main, mutually orthogonal vectors. It turned out that
the fivevectors $\textbf{\textit A}_{\rm eff}$ of all the stars in
the sample can be sufficiently accurately expressed as the linear
combination of three main components (3D-space). Their occurrence
and wavelength behaviour may reflect some astrophysical
significance. Their course with the wavelength is displayed on Fig.
\ref{3main}.

Having the main components defined, we project the vector of
effective amplitude of an mCP into the 3D basis, thus getting three
components $b_1$, $b_2$, $b_3$. They show a specific distribution
with the $B-V$ index.

\begin{figure}
\centerline{\psfig{figure=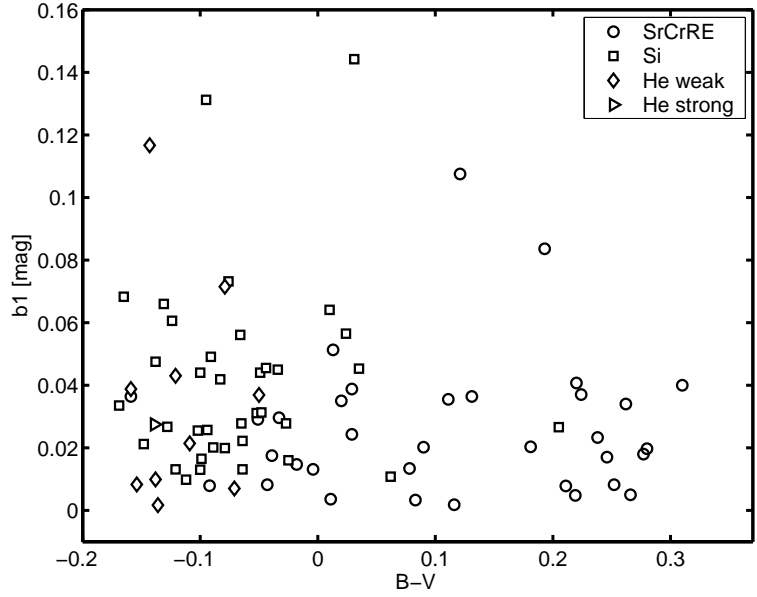,height=8.cm}} \caption{The
dependence of the projection $b_{\rm 1}$ of the vector of the
effective amplitude on the first principal component on the
\emph{B-V} index. All the values are positive suggesting the
photometric spot measured by this quantity looks bright in all
colours studied.} \label{bvb1}
\end{figure}

\begin{figure}
\centerline{\psfig{figure=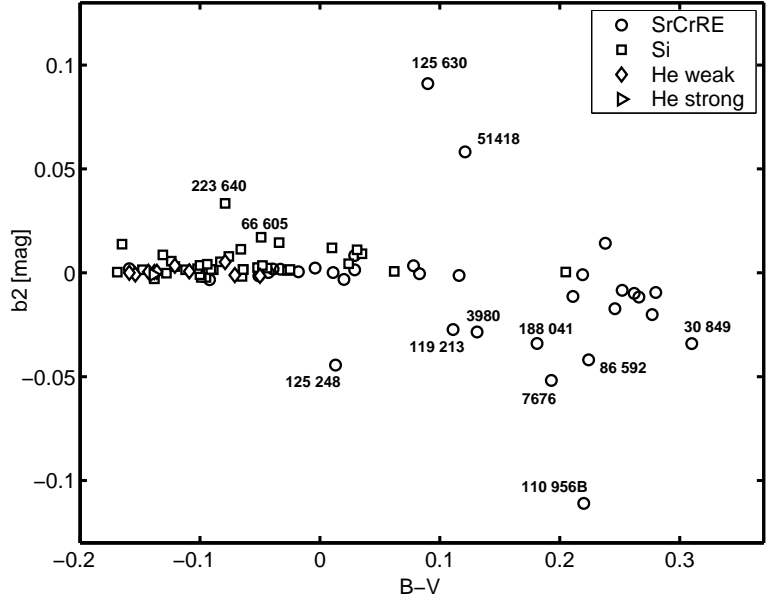,height=8.cm}} \caption{The
dependence of $b_2$ on $B-V$ index. For the hot mCPs the value is
close to zero, for the cooler ones $b_2$ acquires considerable
non-zero values. The corresponding mechanism of energy
redistribution is most prominent in $v$ colour (see
Fig.\ref{3main}), where we see spots in dark. If that spot
interferes with bright spots culminating near the zero phase it
reduces the effective amplitude ($b_2>0$ cf. HD 125603). If the dark
spot is placed on the opposite side of the star, it enlarges the
amplitude ($b_2<0$ cf. HD 110956B).} \label{bvb2}
\end{figure}

\begin{figure}
\centerline{\psfig{figure=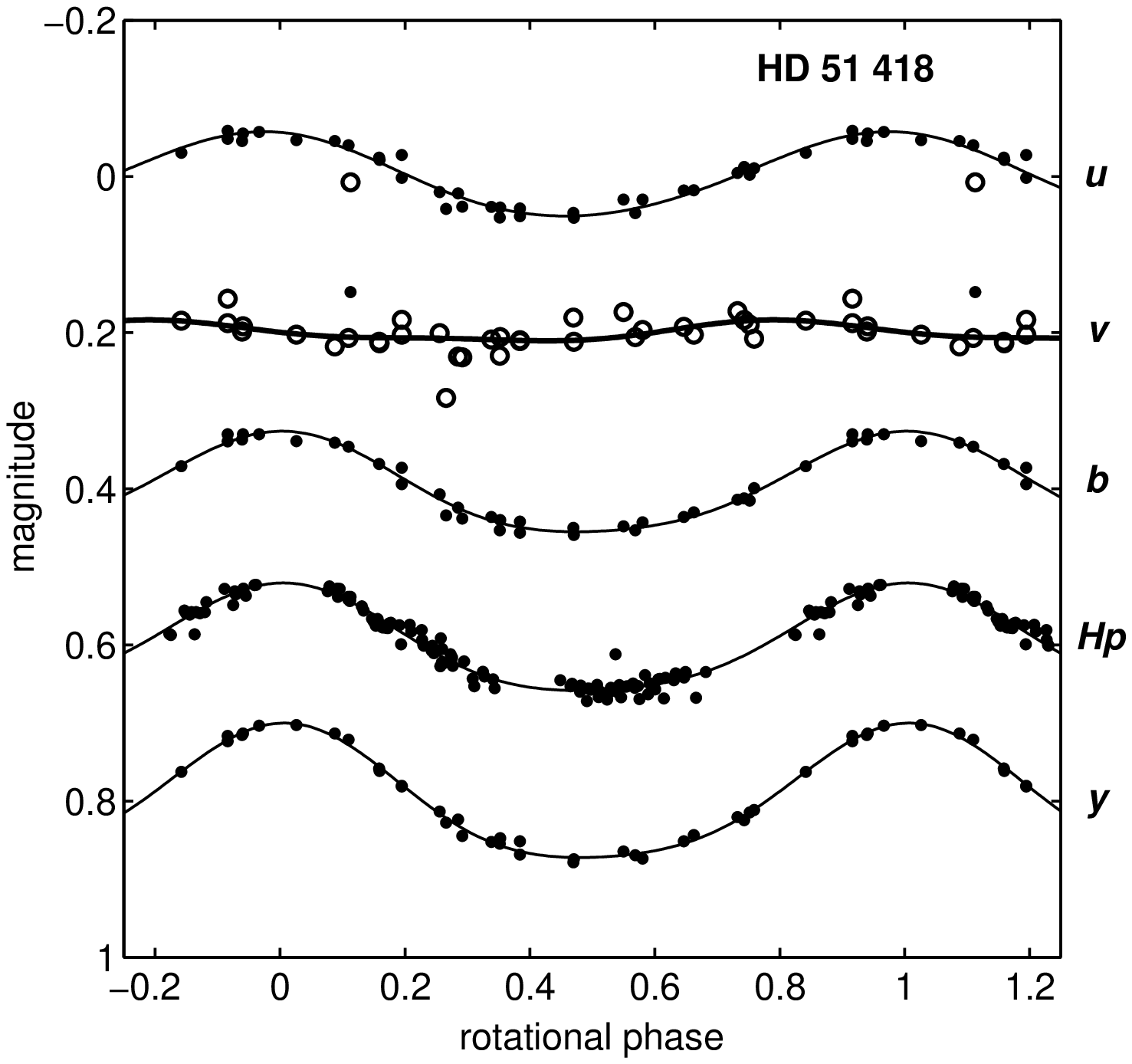,height=7.5cm}
        \psfig{figure=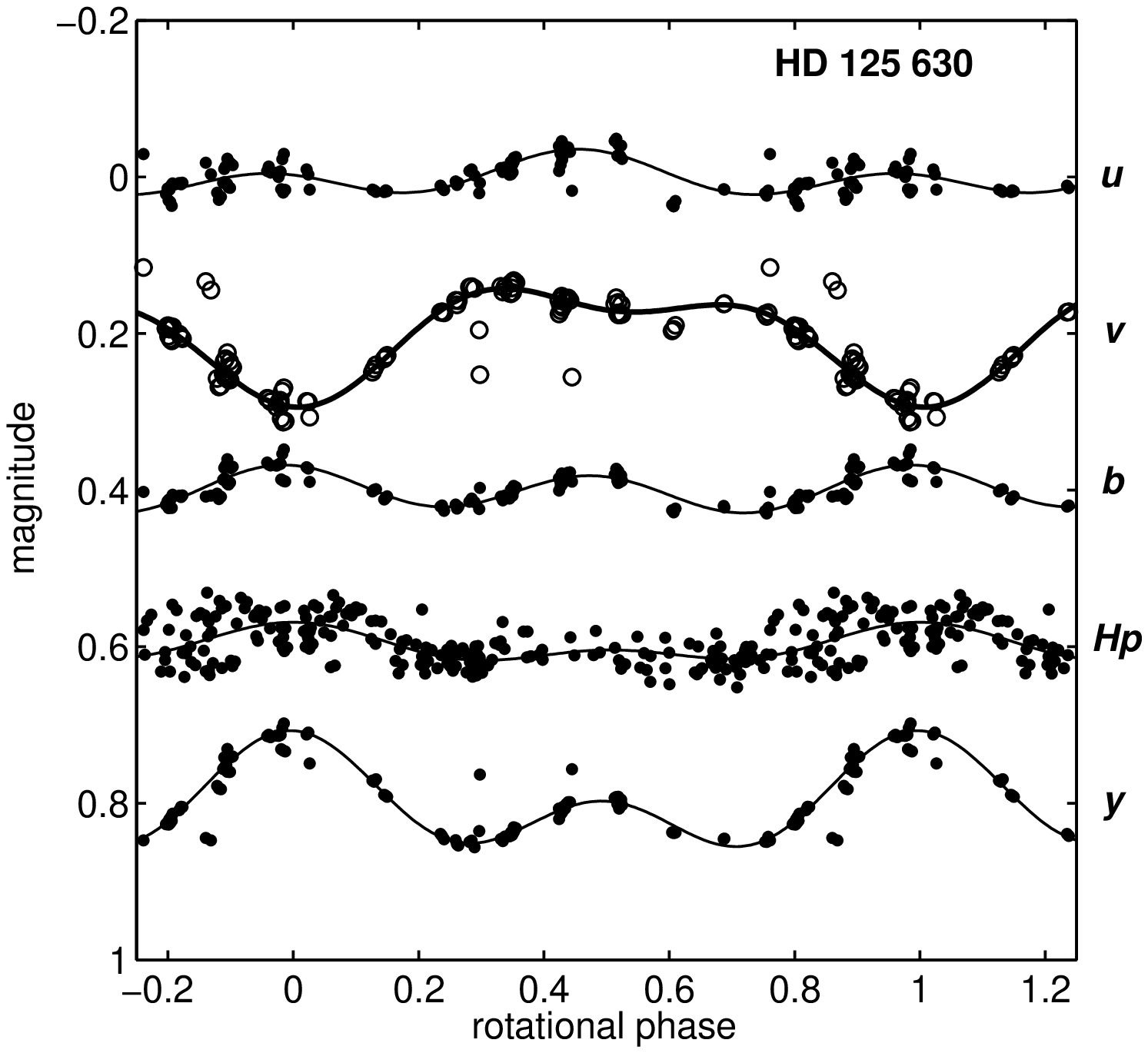,height=7.5cm}}
\caption{The comparison of LCs of two mCP stars with positive $b_2$.
In the case of HD\,51418 the amplitude of LC in $v$ is due to dark
spot suppressed practically to zero, while in the case HD\,125630 we
even see the $v$ LC to be in antiphase with other LCs.}
\label{TwoLC2}
\end{figure}

\begin{figure}
\centerline{\psfig{figure=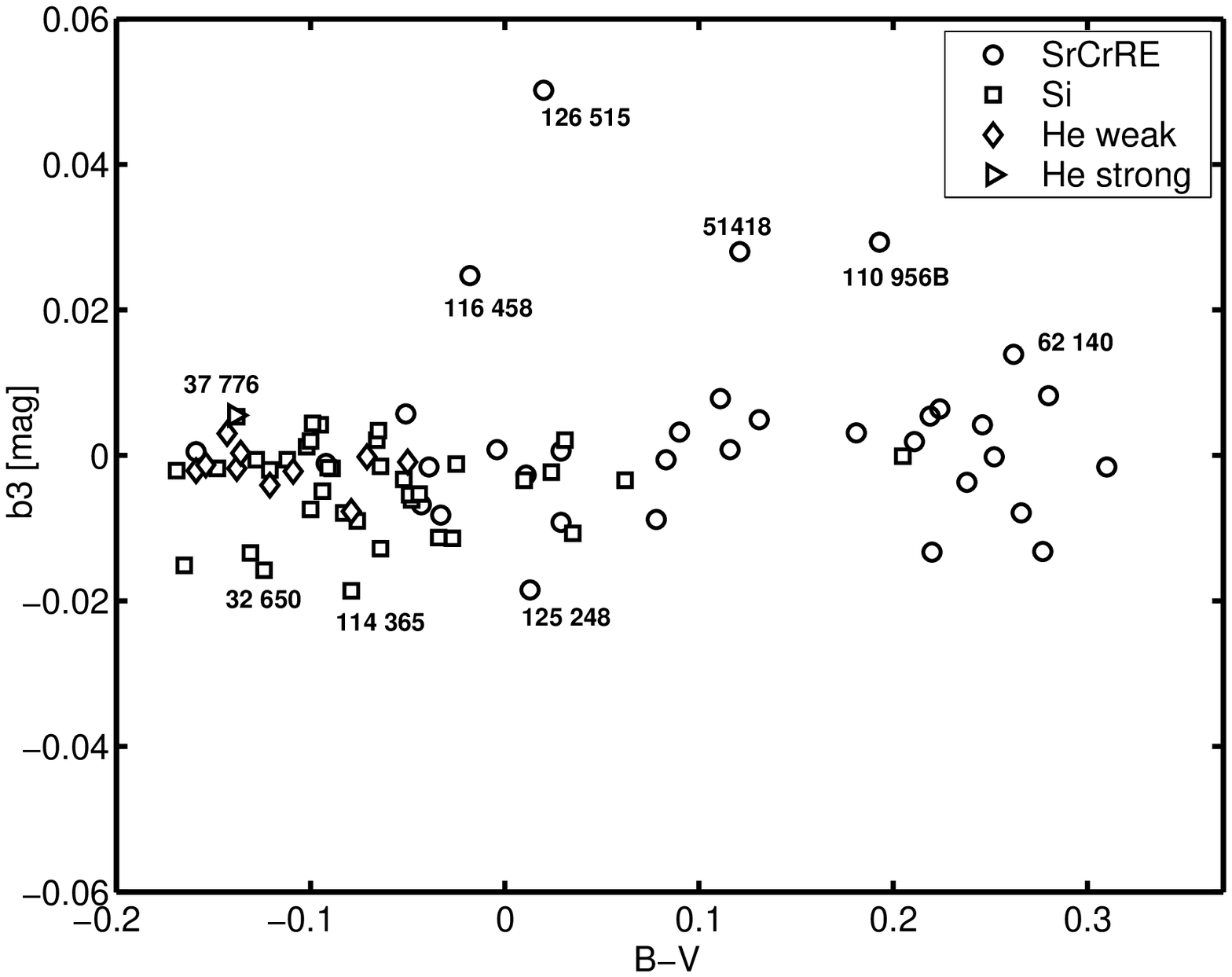,height=8.cm}} \caption{The
dependence of $b_{\rm 3}$ on $B-V$ index. For the hot mCPs the value
is close to zero. For the cooler ones $b_{\rm 2}$ acquires
considerable non-zero values and the mechanism corresponding to this
quantity plays a more significant role. The third principal
component (see Fig.\ref{3main}) acquires large negative value in $u$
implying the spot in this colour is dark. } \label{bvb2}
\end{figure}

\section{Some empirical finding and discussion}
We confirm that the observed light curves of mCP stars can be
satisfactorily well represented by harmonic polynomials of the
second order.
Investigating the wavelength dependence of the
effective amplitude by means of PCA we conclude:
\begin{enumerate}
\item Obviously every of the sample stars has one or more bright spots on
its surface whose contrast decreases with increasing wavelength.
The spots are usually more pronounced on
hotter stars. Nevertheless,
occurrence of these spots is not bound to any of the peculiarity
types.
\item On cooler mCPs, not necessarily of SrCrRE type, one or more spots
looking dark in the $v$ colour occur relatively frequently. These spots are
indicated by the increasing amplitude of the $c_1$-index in the cool
mCPs. We suggest that these spots might be related to the decrease
of Balmer jump in the spectral energy distribution in regions
heavily contaminated by overabundant elements.
\item The spots contrast in the $u$~ may demonstrate themselves either
as bright or
dark, whereas as a rule they are bright on hot and dark on cool
stars.
\end{enumerate}

The brightness contrast of the spots probably is due to the uneven
distribution of chemical elements, horizontal and possibly vertical,
and perhaps also to strong global magnetic field. These dissimilar
atmospheric patches interact differently with the isotropic
radiative flux coming from the interior. Especially, the "colour"
photometric spots may origin due to the different redistribution of
the flux from the UV to visual region caused by the bound-free
transitions and line-blocking. Silicon may play a decisive role in
the light variability, as it is usually overabundant in all subtypes
of the mCP stars.

Consequently, photometric spots are most likely related to the
surface distribution of chemical elements and, possibly, with the
configuration and the intensity of magnetic fields. The chemical
elements use to be distributed in several more or less irregularly
distributed spots (cf. \cite{kvst}), as well as the magnetic field
configuration may differ from an elementary dipole structure
(\cite{kochu}). Thus, a simple pattern of the photometric spots
cannot be expected, as it was documented e.\,g. in the case of
HD\,37776 (\cite{krta}).

The amazing diversity of the forms of the LCs is the result of the
complexity of the photometric spots structure. The study of the
relations between the photometric and spectroscopic spots on the mCP
stars seems to be very promising.

From this point of view, the mCP stars represent a relatively
homogeneous group. However, our conclusions are very preliminary
mainly for they are based only on the analysis of the effective
amplitude. We plan also to take into account the shape of a LC
harbouring information on location of the spots along the surface.

\begin{acknowledgements}
This work was supported by grants GA\,\v{C}R 205/06/0217, VEGA
2/6036/6, and MVTS \v{C}RSR 10/15/2006-7. This research has made use
of NASA's Astrophysics Data System, the SIMBAD database, operated at
CDS, Strasbourg, France and \emph{On-line database of photometric
observations of mCP stars} (\cite{mikdat}).
\end{acknowledgements}


\end{document}